%% file: CVCholography.tex
\documentclass[12pt]{JHEP3}
\pdfoutput=1
\usepackage[english]{babel}
\usepackage{amssymb}
\usepackage{amsfonts}
\usepackage{amsmath}
\usepackage{graphicx}
\usepackage{subfigure}
\usepackage{epsfig}
\usepackage{cite}
\usepackage{psfrag}
\usepackage{slashed}

\usepackage[normalem]%
{ulem}  
\newcommand{\old}[1]{}

\def\erf#1{(\ref{#1})} 
\newcommand{\cA}{{\cal A}}  \newcommand{\cB}{{\cal B}}

\newcommand{\be}{\begin{equation}} \newcommand{\ee}{\end{equation}}
\newcommand{\bea}{\begin{eqnarray}} \newcommand{\eea}{\end{eqnarray}}
\newcommand{\beann}{\begin{eqnarray*}}  \newcommand{\eeann}{\end{eqnarray*}}
\newcommand{\bfig}{\begin{figure}} \newcommand{\efig}{\end{figure}}
\newcommand{\ba}{\begin{array}} \newcommand{\ea}{\end{array}}
\newcommand{\bcen}{\begin{center}} \newcommand{\ecen}{\end{center}}
\newcommand{\btab}{\begin{tabular}} \newcommand{\etab}{\end{tabular}}
\newcommand{\nn}{\nonumber}
\newcommand{\vev}[1]{\left\langle{#1}\right\rangle}

\newcommand{\dd}{{\rm d}}
\newcommand{\e}{{\rm e}}

\newtheorem{Proposition}{Proposition}[section]

\newtheorem{Theorem}{Theorem}[section]
\newtheorem{Lemma}{Lemma}[section]
\newtheorem{Corrolary}{Corrolary}[section]

\newcommand{\bp}{\begin{Proposition}}	\newcommand{\ep}{\end{Proposition}}
\newcommand{\bt}{\begin{Theorem}}	\newcommand{\et}{\end{Theorem}}
\newcommand{\bl}{\begin{Lemma}}		\newcommand{\el}{\end{Lemma}}
\newcommand{\bc}{\begin{Corrolary}}	\newcommand{\ec}{\end{Corrolary}}

\title{Anomalous transport coefficients from Kubo formulas in Holography
}
\author{ Irene Amado$^{a}$, Karl Landsteiner$^{b}$,  Francisco Pena-Benitez$^{b}$\\
$^{a}$SISSA\\
Via Bonomea, 265\\
I-34136  Trieste, Italy\\
E-mail: \email{iamado@sissa.it}\\
\\
$^{b}$Instituto de F\'\i{}sica Te\' orica CSIC-UAM\\
c/ Nicol\' as Cabrera 13-15\\
Universidad Aut\' onoma de Madrid\\
E-28049  Madrid, Spain\\
E-mail: \email{karl.landsteiner@csic.es, fran.penna@uam.es}
}

\preprint{IFT-UAM/CSIC-11-07 \\ SISSA 08/2011/EP \\ ESI 2305 }

\abstract{In the presence of dense matter quantum anomalies give rise to two new transport phenomena. An anomalous current is generated either by an external magnetic field or through vortices in the fluid carrying the anomalous charge. The associated transport coefficients are the anomalous magnetic and vortical conductivities. Whereas a Kubo formula for the anomalous magnetic conductivity is well known we develop a new Kubo type formula that allows the calculation of the vortical conductivity through a two point function of the anomalous current and the energy current. We also point out that the anomalous vortical conductivity can be understood as a response to a gravitomagnetic field. We apply these Kubo formulas to a simple holographic system, the R-charged black hole.}

\begin{document}

\input{intro}

\input{kubo-formula}

\input{kubo-hologra}

\input{conclusion}

\appendix

\input{second-order-sol}

\input{anal-sol}

\input{v-diff-0}

\begin{acknowledgments}
We would like to thank 
Johanna Erdmenger,  Antti Gynther, Matthias Kaminski, Ingo Kirsch, Anton Rebhan, Andreas Schmitt,
Laurence Yaffe and Ho-Ung Yee, for useful discussion. 
We also would like to thank the Erwin Schr\"odinger Institute Vienna for hospitality during
the workshop on AdS Holography and Quark Gluon Plasma. 
This work has been supported by Accion Integrada Hispanoaustriaca HA2008-0003, 
Plan Nacional de Alta Energ\'\i as FPA2009-07908, 
Comunidad de Madrid HEP-HACOS S2009/ESP-1473.
\end{acknowledgments}

\bibliographystyle{JHEP}
\bibliography{CVCholography}

\end{document}

%% file: intro.tex
\section{Introduction}
Anomaly induced charge separation effects in heavy ion collisions are currently under intense discussion \cite{cpodd:2010}.
There are basically two such effects called the chiral magnetic effect and the chiral vortical effect. 
Both effects rely essentially on the presence of an imbalance between left-handed and right-handed quarks. In many
models this imbalance is described by introducing a chemical potential for the axial charge.
The chiral magnetic effect describes the generation of a current in a background magnetic field and the
chiral vortical effect describes the generation of  a current in the presence of a vorticity field in the charged
fluid. 

The chiral magnetic effect has actually a somewhat complicated history as it seemingly has been discovered independently 
several times \cite{Alekseev:1998ds, Giovannini:1997eg}. In any case its relevance to heavy ion collisions has been argued for in \cite{Kharzeev:2007jp} and from then on there has
been an intensified activity in deriving and understanding it from various perspectives. Most of the work has been done from
a traditional field theory approach \cite{Fukushima:2008xe, Kharzeev:2009pj, Fukushima:2010vw, Fukushima:2009ft, 
Muller:2010jd, Orlovsky:2010ga, Pu:2010as, Nam:2010nk, Fu:2010rs, Asakawa:2010bu, Fukushima:2010fe, Fu:2010pv, Nam:2009hq, Kharzeev:2009fn }. Since it is related to the anomaly one would naturally believe that the chiral magnetic conductivity is an exact result and should apply also at strong coupling. Indeed it has recently been shown that higher loop
corrections do not contribute and thus the result is indeed exact in perturbation theory \cite{Hong:2010hi}. Non-anomalous corrections to it due to interactions have been however argued for in \cite{Fukushima:2010zz}.
Lattice calculations of the chiral magnetic effect have been reported in \cite{Buividovich:2009wi,Abramczyk:2009gb,Buividovich:2009zj}. 
Recently it also has been shown that at finite magnetic field there are now propagating hydrodynamic modes called chiral magnetic waves
\cite{Kharzeev:2010gd}.

Results reported in holographic models were however consistent with a vanishing chiral magnetic conductivity or used the consistent form of the anomaly with a non-conserved vector current\cite{Rebhan:2009vc, Yee:2009vw,Rebhan:2010ax} \footnote{See also \cite{Gorsky:2010xu}.}.   This discrepancy
in the holographic calculations has been resolved in \cite{Gynther:2010ed}. There it was pointed out that a chemical potential can be introduced in
various gauge equivalent forms. Since gauge invariance is however violated by the anomaly these different formalism have
to be kept distinguishable during the calculation and it was shown that the weak coupling result corresponds to a formulation
in which the chemical potential is not represented by a background value for the temporal component of a fiducial gauge field
but rather by certain boundary conditions on the fields. In fact this seems the most natural way of introducing  a chemical
potential for an anomalous gauge symmetry, since after all vector fields that couple to anomalous currents are absent in nature.\footnote{In \cite{Rubakov:2010qi} it was argued that a chemical potential should be introduced only for non-anomalous charges. By addition of
a Chern-Simons current a truly conserved axial charge can indeed be defined. This charge is however only gauge invariant if
it is integrated over all of space. Therefore this approach seems not directly applicable to situations where the charge is supposed
to be localized inside the small fireball of heavy ion collision. We rather prefer to work with chemical potentials for anomalous charges that are not truly conserved but localizable within small regions in a gauge invariant way.}

The anomalous vortical effect on the other hand has first been noticed in a holographic calculation of the hydrodynamics of 
R-charged asymptotically anti de Sitter black holes \cite{Erdmenger:2008rm, Banerjee:2008th}. Soon it was pointed out that such a term in the constitutive relations for an anomalous current is not only allowed but even necessary in order to obtain an entropy current whose divergence is positive definite\cite{Son:2009tf}. Recently it was suggested that the interplay of chiral magnetic effect and chiral vortical effect leads to interesting signatures of baryon number separation and charge separation in heavy ion collisions \cite{Kharzeev:2010gr} whereas \cite{KerenZur:2010zw} argued for
enhanced production of spin-excited hadrons. The conserved electromagnetic current 
has been explicitly incorporated in the hydrodynamic constitutive relations recently on \cite{Sadofyev:2010pr} whereas \cite{Torabian:2009qk,Neiman:2010zi}
considered an extension with global or gauged abelian and non-abelian symmetries. Very recently a 
STU model realizing the chiral magnetic and the chiral vortical effects has been presented in \cite{Kalaydzhyan:2011vx}.

Let us briefly discuss how  a chemical potential can be introduced in field theory. Often it is viewed as a deformation of
the theory by a new coupling of the form $H \rightarrow H -\mu Q$ where $H$ denotes the Hamiltonian, $\mu$ the chemical
potential and $Q$ the charge operator. If $Q$ is the zero component of a conserved current $j^\mu$ then we can think of this as
arising from coupling the current to a (fiducial) gauge field $A_\mu$  and choosing a constant background value for the temporal
component of the gauge field $A_0 = \mu$. In a finite temperature context fields obey the KMS state condition
$\phi(t+i/T) = \eta \phi(t)$, where $\eta=\pm 1$ for either Bosons or Fermions. Due to the underlying gauge invariance we can
however also introduce the chemical potential in a different but equivalent way. We can perform a gauge transformation that sets
$A_0=0$. That does of course not mean that the chemical potential has vanished from our theory. Since the (imaginary) time direction is
compactified due to the KMS condition such a large gauge transformation changes the boundary conditions on the fields.
In fact the correct KMS state conditions in the gauge with $A_0=0$ are 
\begin{equation}
\phi(t+i/T) = \eta\, e^{-q\mu/T} \phi(t)\,,
\end{equation}
where $q$ is the charge of the field $\phi$. It seems useful to think of the first formulation as a deformation of the dynamics of the 
system whereas the second formulation leaves the dynamics, i.e. the Hamiltonian $H$ unchanged but modifies the state 
space (see e.g. \cite{Landsman:1986uw, Evans:1995yz, Kogut:2000ek}). 

If we try now to define an analogue of  the chemical potential in the case of an anomalous theory we have to be aware of the fact
that these two formulations  might give different results since a gauge transformation on the fields is not anymore an invariance
of the quantum theory. Following \cite{Gynther:2010ed} our approach in holography is to choose a formulation that distinguishes between the elementary definition of the chemical potential as the energy that is needed to introduce a unit of charge into the system from
the actual value of the temporal gauge field. In the holographic setup the energy needed to introduce a unit charge into the system is given by the potential difference between the horizon and the boundary. This quantity does not change under gauge
transformations even in the anomalous case. This condition describes the state of the system.
 On the other hand we can then allow for an arbitrary gauge field background that is not a priori related to the chemical potential. 
 Choosing such a non-vanishing value for $A_0$ will then also deform the dynamics. In \cite{Gynther:2010ed} it was shown that calculating
 the chiral magnetic conductivity with the deformed dynamics leads to a zero result but that a non zero result coinciding 
with the weak coupling result \cite{Kharzeev:2009pj} is obtained if this deformation is switch off.

Whereas the actual chiral magnetic effect relevant to the physics of heavy ion collision depends on the interplay of a conserved
vector current and a non-conserved chiral $U(1)$ current we will focus in this paper on the most simple setup where only one
anomalous $U(1)$ current is present. 
One of the main results of this paper is a Kubo formula for the anomalous magnetic and vortical
conductivities $\xi_{B}$,  $\xi_{V}$ as they are defined through the constitutive relations in \cite{Erdmenger:2008rm, Banerjee:2008th,Son:2009tf}. 
These Kubo type formulas are:

\begin{eqnarray}
 \label{eq:cmc}\xi_{ B}  &=& \lim_{k_n\rightarrow 0} \frac{-i}{2 k_n} \sum_{k.l} \epsilon_{nkl} \left(\left. \left\langle J^{k} J^{l} \right\rangle - 
\frac{n}{\epsilon+P}\left\langle T^{tk} J^{l} \right\rangle \right)\right|_{\omega=0, A_0=0}\,,\\
\label{eq:cvc}\xi_{V} &=& \lim_{k_n\rightarrow 0} \frac{-i}{2 k_n} \sum_{k.l} \epsilon_{nkl} \left.\left( \left\langle J^k T^{tl} \right\rangle - 
\frac{n}{\epsilon+P}\left\langle T^{tk} T^{tl} \right\rangle \right) \right|_{\omega=0}\,.
\end{eqnarray}
Here $J^k$ are the spatial components of the anomalous current, $T^{tk}$ is the energy flux and $k_n$ the momentum.
The correlators are retarded ones. 

That we have to compute the sum of two correlators has to do with the particular way the coefficients are defined in the 
constitutive relation and is related to an ambiguity in the definition of the fluid velocity at first order in derivatives. We also have explicitly included in our definition of the anomalous magnetic conductivity that the temporal component of the gauge has to vanish. It is remarkable that the anomalous vortical conductivity does
not depend on the gauge field background and therefore can be evaluated for any $A_0$. 

This paper is organized as follows. In section two we derive the Kubo formulas for the anomalous conductivities from the constitutive relations derived in \cite{Son:2009tf}. We point to an inherent ambiguity in the definition of the fluid velocity. Furthermore we define new elementary conductivities relying on the interpretation of the anomalous vortical effect as a gravitomagnetic effect and relate them with the Landau frame conductivities defined above.

We then go on and define our holographic model, which is nothing but (a truncation of) the asymptotically R-charged black hole background. 
We define the current and show
that it obeys an anomalous conservation law. We compute then the relevant matrix of retarded Green functions analytically
in the hydrodynamic limit and show that applying the Kubo formulas (\ref{eq:cmc}), (\ref{eq:cvc}) the  results established in 
\cite{Son:2009tf} are obtained. We find that the constitutive relations for the anomalous current have to be modified by including a Chern-Simons coupling.  
We would also like to comment that the computation of the retarded correlators for the holographic model considered in this work has been addressed before in \cite{Matsuo:2009xn}, though the ambiguity in the 
fluid velocity was not addressed and only the magnetic conductivity was computed. In fact, the Green functions found by the authors correspond to the particular case of setting that velocity to zero, i. e. neglecting the energy flow 
due to the external background magnetic and vorticity fields\footnote{Let us also note that chiral dispersion relations in the shear sector
have been found in \cite{Sahoo:2009yq}. This is however a higher order effect and not captured by first order hydrodyamics which is our prime interest here.}. 
We then proceed and compute frequency dependent conductivities numerically. 

In section four we draw our conclusions and discuss possible further applications of the Kubo formalism. In particular we
emphasize that the Kubo formula allows for a first principle calculation of the chiral vortical conductivity along the lines of
the calculation of the chiral magnetic conductivity of \cite{Kharzeev:2009pj,progress}.

In Appendix \ref{sec:app2nd} and \ref{solw0} we collect the lengthy expressions for the perturbed action and the analytic solutions for the fluctuations in the hydrodynamic regime. 
In Appendix \ref{vdiff0} we explore the dependence of the correlators and the conductivities in the fluid velocity at zero frequency.

%% file: kubo-formula.tex
\section{Kubo formulas and constitutive relations}

Hydrodynamics can be understood as an effective, dissipative field theory for a fluid. As is usual for effective field theories 
it is organized as a derivative expansion. To lowest order one starts with the equilibrium expressions of energy momentum
tensor and currents. Dissipative effects appear at first order in derivatives. According to the underlying space time symmetries
the derivative terms can be organized in certain tensor structures. The most important input parameters to hydrodynamics
are then the coefficients in front of these derivative tensor structures. These are the transport coefficients, such as viscosities
or conductivities. These constitutive relations can also be written down in the presence of generic external fields, such as
electric and magnetic fields or gravitational fields. 

The constitutive relations for a relativistic fluid carrying charge of an anomalous $U(1)$ symmetry have first been obtained
via holographic methods in \cite{Erdmenger:2008rm,Banerjee:2008th}. They have been further studied from a purely
hydrodynamic point of view in \cite{Son:2009tf}. These constitutive relations are
 \begin{eqnarray}
 T^{\mu\nu} &=& (\epsilon + P) u^\mu u^\nu + P g^{\mu\nu}  + \tau^{\mu\nu} \,,\\
 J^\mu &=& n u^\mu + \nu^\nu\,.
 \end{eqnarray}
Here $\epsilon$ is the energy density, $P$ the pressure, $n$ the charge density and $u^\mu$ the local fluid velocity.
The fluid velocity can be chosen such that the dissipative parts $\tau^{\mu\nu}$ and $\nu^\nu$  obey $u_\mu \tau^{\mu\nu}=0$ and  $ u_\mu \nu^\mu = 0$. 

In the presence of an external background gauge field and assuming the current to be anomalous the conservation equations take
the form
\begin{eqnarray}
\nabla_\mu T^{\mu\nu} &=& F^{\nu\mu} J_\mu\,,\\
\nabla_\mu J^{\mu} &=& c \,E_{\mu} B^{\mu}\,.
\end{eqnarray}

The constitutive relations for the dissipative parts take then the form \cite{Son:2009tf}
\begin{eqnarray}
\tau^{\mu\nu} &=& -\eta P^{\mu\alpha} P^{\nu\beta} \left( \nabla_\alpha u_\beta +\nabla_\beta u_\alpha - \frac 2 3 \nabla_\lambda u^\lambda g_{\alpha\beta} \right) - \zeta P^{\mu\nu} \nabla_\lambda u^\lambda\,, \\
\nu^\mu &=& -\sigma T P^{\mu\nu}\nabla_\nu\left(\frac{\mu}{T}\right) + \sigma E^\mu + \xi_{B} B^\mu + \xi_{V} \omega^\mu \, ,
\end{eqnarray}
where $P^{\mu\nu}=g^{\mu\nu}+u^\mu u^\nu$, $E^\mu = F^{\mu\nu}u_\nu$, $B^\mu = \frac 1 2 \epsilon^{\mu\nu\rho\lambda}u_\nu F_{\rho\lambda}$ and
$\omega^\mu = \epsilon^{\mu\nu\rho\lambda}u_\nu\partial_\rho u_\lambda$ is the vorticity of the fluid. 
In addition to the usual transport coefficients, the shear viscosity $\eta$, the bulk viscosity $\zeta$ and the conductivity $\sigma$
there are two new, parity violating terms proportional to the magnetic field $\xi_B$ and the fluid vorticity $\xi_V$. These are induced by the anomaly. 

We now want to derive Kubo formulas for the anomalous transport coefficients. First let us assume that the fluid is close to rest $u^\mu=(1,v^x,0,v^z)$ with only small velocities such that $u^\mu u_\mu=-1+O(2)$.  If we interpret the constitutive relations as the one point functions of the current and the energy momentum tensor in the presence of external sources, we can obtain the two point function of currents by differentiating with respect to the gauge potential and the metric\footnote{Including a fiducial metric in the constitutive relations of hydrodynamics for the purpose of deriving 
Green-Kubo-Mori formulas goes back to Luttinger's theory thermal transport coefficients \cite{Luttinger:1964zz}, where he states that
if the gravitational field didn't exist, one could invent it for the purpose of deriving the transport coefficients.}.

In the constitutive relations there are no
terms that depend on the metric only. This is a consequence of the fact that covariant curvature expressions are second order in derivatives and
therefore can not appear as long as we restrict ourselves to first order hydrodynamics. Coupling to the metric is achieved simply by taking all derivatives as covariant ones. 

We now assume that in addition to the velocity field we have a metric perturbation of the form $g_{\mu\nu} = \eta_{\mu\nu} +h_{\mu\nu}$ with only $h_{tx}$, $h_{tz}$, $h_{yx}$ and $h_{yz}$ different from zero. The gauge field perturbations we take to be 
$A_\mu = (0,a_x,0,a_z)$. All unknown functions are supposed to depend only on $(t,y)$. To first order in derivatives, sources and the velocity the relevant constitutive relations are now
\begin{eqnarray}
J^m &=& n v^m - \epsilon^{mn}\xi_V \left( \partial_y h_{tn} + \partial_y v^n\right) - \sigma \partial_t a_m - \xi_B \epsilon^{mn} \partial_y a_n  + O(2)\,, \label{eq:constitituve_J}\\
T^{tm} &=& (\epsilon + P) v^m + P h_{tm} \label{eq:constitutive_Tt} + O(2)\,,\\
T^{ym} &=& -P h_{ym} - \eta ( \partial_y v^m +\partial_t h_{ym}) \label{eq:constitutive_Tx} + O(2) \,,
\end{eqnarray}
with $m \in \{ x,z \}$. In addition we have the (non-)conservation law
\begin{equation}
[(\epsilon +P) \partial_t - \eta \partial^2_y] v^m + n \partial_t a_m + (\epsilon+P)\partial_t h_{tm} - \eta \partial_{t}\partial_{y} h_{ym} =O(3)\,.
\end{equation}
The orders on the right indicate that these equations hold up to $n$-derivatives. We could now eliminate the velocity from the 
conservation equation of the stress tensor. However, this equation does have a non-vanishing zero mode to first order in the
derivatives. For simplicity let us take the zero frequency sector. We see then that any velocity of order $O(\partial_y)$ solves
the conservation equation. Even at non-zero frequency we could still have a velocity of the form 
$v_m\sim \frac{\partial_y^3}{(e+P)\partial_t - \eta \partial_y^2} f(t,y)$ for an arbitrary function $f(t,y)$. In non-anomalous theories
we can exclude such terms since parity conservation together with the $SO(3)$ rotational symmetry would tell us that the spatial momentum 
can enter only as $k^2$ in the
equations. In the presence of the parity violating anomaly we can not assume this anymore and have to take the kernel of the
energy-momentum conservation into account. We will denote this kernel as $v_m^{(1)}$ and treat it as an arbitrary function of
the sources. 

Let us further simplify to zero frequency. We find then
\begin{eqnarray}
T^{tm} &=& (\epsilon+P) v_m^{(1)} + P h_{tm}\,,\\
J^m &=& n v_m^{(1)} -  \epsilon^{mn} \left( \xi_V \partial_y h_{tn} + \xi_B\partial_y a_n \right)\,.
\end{eqnarray}
We can differentiate now with respect to $h_{tm}$ or $a_n$ and eliminate the undetermined $\left\langle  v_m^{(1)} T^{tn}\right\rangle$ and $\left\langle  v_m^{(1)} J^{n}\right\rangle$ correlators and arrive at
\begin{eqnarray}
\left.\left(\left\langle J^m J^n \right \rangle - \frac{n}{\epsilon+P} \left\langle T^{tm} J^n \right \rangle\right)\right|_{\omega=0} &=& \xi_B \epsilon^{mnl} i k_l +O(k^2)\,,\\
\left.\left(\left\langle J^m T^{tn} \right \rangle - \frac{n}{\epsilon+P} \left\langle T^{tm} T^{tn} \right \rangle\right)\right|_{\omega=0} &=& -\frac{n P}{\epsilon+P} \delta^{mn} + \xi_V \epsilon^{mnl} i k_l +O(k^2)\,.
\end{eqnarray}
From this the Kubo formulas \erf{eq:cvc} and \erf{eq:cmc} follow immediately. 

In fact even without using the (non-)conservation equation for the energy momentum tensor we can find some useful
relationships between Green functions. We can solve the constitutive relation (\ref{eq:constitutive_Tt}) for $v^m$ and
plug it into the one for the current. Upon differentiating with respect to the metric component $h_{tn}$ or the gauge
field $a_n$ we find that for $m=x$, $n=z$ the following relations hold
\begin{eqnarray}
\label{xib}\hspace{-0.5cm}\frac{i}{k_y} \left[\langle J^x J^z\rangle -\frac{n}{\epsilon+P} \langle T^{tx} J^z\rangle\right] &=& \xi_B + \frac{\xi_V}{\epsilon+P} \langle T^{tz} J^z\rangle + O(k) \,,\\
\label{xiv}\hspace{-0.5cm}\frac{i}{k_y} \frac{(\epsilon+P)}{\epsilon+\langle T^{tz}T^{tz}  \rangle}\left[ \langle J^x T^{tz}\rangle -\frac{n}{\epsilon+P} \langle T^{tx} T^{tz}\rangle \right]&=& \xi_V +O(k)\,.
\end{eqnarray}
To lowest order in $\omega$ and $k$ we have $\langle T^{tz} T^{tz}\rangle =P$ and $ \langle T^{tz} J^z\rangle=0$
and we recover the previous Kubo type formulas.

Finally we need to understand why the zero-mode velocity $v^{(1)}_m$ is present. We know that a variation in the
charge distribution will cost us an energy of the form $\delta \epsilon = \mu \delta Q$. If we imagine a test charge
$\delta Q$ moving through the charged plasma it will therefore generate a current $\delta \vec{J}$ and induce also an energy
current of the form $\delta T^{ti} = \mu \delta J^i$. For a finite current we should integrate this over $\mu$
and obtain $T^{ti} = \int_0^\mu \mu'  \frac{\dd J^i}{\dd\mu'} d\mu'$. 
In the Landau frame this implies however the generation of a velocity through $T^{ti} = (\epsilon +P) v^i$. 
If we are interested in the charge current that sits on top of the one produced by the fluid velocity we should subtract the component of the current that is due to the drag of the fluid in the Landau frame 
\begin{equation}
J^i_{drag} = n v^i =  \frac{n}{\epsilon+P} T^{ti}\,. 
\end{equation}

From what we outlined above it follows then that the transport coefficients $\xi_B$ and $\xi_V$ that appear in the Landau frame constitutive relations are\footnote{We would also like to point to \cite{Sadofyev:2010is} where using an effective
field theory approach it has been argued that the terms of higher order in the chemical potential of the anomalous conductivities are
ambiguous due to infrared effects.}
\begin{eqnarray}
\xi_B &=& = \sigma_B - \frac{n}{\epsilon+P} \int_0^\mu \mu' \dd \sigma_B(\mu') \,,\\
\xi_V &=& = \sigma_V - \frac{n}{\epsilon+P} \int_0^\mu \mu' \dd \sigma_V(\mu') \,.
\end{eqnarray}
Using these relations we find the following form of the relevant correlators
\begin{eqnarray}
\label{sigmab}\left \langle J^m J^n \right\rangle &=& i  \epsilon^{mnl} k_l  \; \sigma_B\, ,\\
\left \langle T^{tm} J^n \right \rangle &=& i  \epsilon^{mnl} k_l \int_0^\mu \mu' \dd\sigma_B(\mu')  \, \label{eq:integrali},\\
\left\langle T^{tm} T^{tn} \right\rangle &=& i \epsilon^{mnl} k_l  \int_0^\mu \mu'^2 \dd\sigma_B(\mu') \, \label{eq:integralii} .
\end{eqnarray}

The second line in these equations fixes also the chiral vortical conductivity by complex conjugation 
\be\label{sigmav}\langle J^m T^{tn}\rangle = \langle  T^{tn} J^m\rangle^\dagger = i \epsilon^{mnl} k_l \sigma_V.\ee

We also would like to point out now that the vortical effect can be understood as a anomalous gravitomagnetic
effect. For slowly rotating and well localized matter distributions it is often convenient to  write the
gravitational field as
\begin{equation}
\dd s^2 = -(1-2 \Phi_g) \dd t^2 - 2 \vec{A}_g \dd\vec{x} \dd t + (1+2\Phi_g)\delta_{ij}\dd x^i \dd x^j\,.
\end{equation}
$\Phi_g$ is the Newtonian gravitational potential whereas $\vec{A}_g$ is the gravitomagnetic potential. 
In full analogy  we can define the new elementary chiral magnetic and vortical conductivity through 
\begin{eqnarray}
\vec{J} &=& \sigma_B \vec{B}\, ,\\
\vec{J} &=&  \sigma_V \vec{B_g}\, ,
\end{eqnarray}
where we introduced a ``gravitomagnetic'' field $ B_g^i = \epsilon^{ijk} \partial_j A_{g,k}$ for the  metric 
fluctuation\footnote{See \cite{Mashhoon:2003ax} for a review on the gravitoelectromagnetic approximation to 
general relativity. The gravitoelectric field $\vec{E}_g = -\vec\nabla \Phi_g$ has of course also a nontrivial 
effect and measures the (thermo)electric conductivity which is most easily seen when we substitute the temperature
in the constitutive relations by the gravitational potential dependent local Tolman temperature \cite{Luttinger:1964zz}.}.

We will now solve for the retarded two--point functions in the holographic model. We do this analytically in the hydrodynamic 
limit. Here we will find an analogous ambiguity that appears as an undetermined integration constant and that is related with the arbitrariness in the definition of the fluid 
velocity already mentioned. This ambiguity disappears once the correct boundary conditions are imposed. Numerically
we can work to all orders in frequency and momentum and there we will find that all correlators are uniquely determined.

%% file: kubo-hologra.tex
\section{Kubo formulas in Holography}\label{sec:holo}

The holographic dual description of a $4$-dimensional $U(1)$ chiral gauge theory is simply 
given by a $5$-dimensional Einstein-Maxwell model supplemented with a topological term \footnote{For the sake of clarity let us mention that here we are defining the $\epsilon$ tensor to be related with the 
Levi-Civita symbol by $\epsilon _{ABCDE}=\sqrt{-g}\,\epsilon (ABCDE)$, with $\epsilon (0r123) =\epsilon (0123)= -1 $. This normalization is consistent with having positive Chern-Simons parameter $\kappa$.},
\be \label{eq:bulkaction}
S= \frac{1}{16 \pi G} \int \dd ^5 x \sqrt{-g} \left( R+\frac{12}{L^2}-\frac{1}{4}F^2+\frac{\kappa}{3} \epsilon ^{MABCD} A_M F_{AB} F_{CD} \right) \,,
\ee
where we can keep the value of the Chern-Simons parameter $\kappa$ to be arbitrary. Latin indices denote bulk coordinates whereas Greek indices denote boundary coordinates. The equations of motion 
\bea \label{eq:EOMs}
0 &=& R_{AB} -\frac{1}{2} \left( R+\frac{12}{L^2}\right) g_{AB} + \frac{1}{2} F_{AC} F^C\,_B + \frac{1}{8} F^2 g_{AB}\,, \nn \\
0 &=& \nabla _B F^{BA} + \kappa \epsilon ^{AMNCD} F_{MN} F_{CD} \,,
\eea
admit the following exact AdS Reissner-Nordstr\"om black-brane solution 
\bea
\dd s^2&=& \frac{r^2}{L^2}\left(-f(r) \dd t^2 +\dd \vec{x}^2\right)+\frac{L^2}{r^2 f(r)}  \dd r^2\,,\nn\\
A&=&\phi(r)\dd t = (\beta-\frac{\mu \,r_{{\rm H}}^2}{r^2})\dd t\,,
\eea
where the horizon of the black hole is located at $r=r_{\rm H}$ and the blackening factor of the metric is
\be
f(r)=1-\frac{M L^2}{r^4}+\frac{Q^2 L^2}{r^6}\,.
\ee
The parameters $M$ and $Q$ of the RN black hole are related to the chemical potential $\mu$ and the horizon $r_H$ by
\be
M=\frac{r_{\rm H}^4}{L^2}+\frac{Q^2}{r_{\rm H}^2}\quad,\quad Q=\frac{\mu\, r_{\rm H}^2}{\sqrt{3}}\,.
\ee
The Hawking temperature is given in terms of these black hole parameters as
\be
T=\frac{r_{\rm H}^2}{4\pi\, L^2} f(r_{\rm H})' = \frac{ \left(2\, r_{\rm H}^2\, M - 3\, Q^2 \right)}{2 \pi \,r_{\rm H}^5} \,.
\ee
The pressure of the dual gauge theory is
\begin{equation}
P = \frac{M}{16\pi G L^3}\,.
\end{equation} 
The theory is conformal obeying $\epsilon = 3 P$.

Accordingly to the discussion in the introduction, we identify $\mu$ as the chemical potential of the gauge theory since it is precisely the gauge invariant 
quantity corresponding to the energy needed to introduce a unit charge into the system, i. e. it is the difference between the scalar potential at the boundary 
and at the horizon. On the other hand, the parameter $\beta$ corresponds to the boundary value of $A_0$ and as already mentioned, it can be viewed as a deformation 
of the dynamics of the system. Following the arguments in \cite{Gynther:2010ed}, it will be eventually set to zero in order to recover the weak coupling result 
for the chiral magnetic conductivity, but for the moment we keep it as an arbitrary gauge background for the boundary theory.

The study of the effect of the anomaly in the hydrodynamic behavior of the chiral gauge theory and the extraction of the corresponding transport coefficients 
passes through computing the holographic two--point correlation functions of the current and the stress tensor. For doing so, we expand the action to second order in perturbations 
of both the gauge field and the metric,
\be 
A_M\rightarrow A_M + a_M\,;\quad g_{AB} \rightarrow g_{AB} + h_{AB}\,. 
\ee 

As a first step, we use the action to first order in perturbations to define the current coupled to the gauge field and the stress-energy tensor of the dual field 
theory. To first order in gauge fluctuations, the action on-shell reduces to a boundary term
\be
\delta S^{(1)} = \frac{1}{16\pi G} \int \dd ^4 x \left\lbrace\sqrt{-g} \left(  F^{Ar} + \frac{4\kappa}{3} \epsilon ^{rABCD} A_{B} F_{CD} \right) a_A \right\rbrace\Big{|}_{r\to\infty}\,,
\ee
from which we can read off the boundary current, that applying the holographic dictionary is given by
\be 
J^{\mu}=\frac{\delta S}{\delta A_{\mu}(r\to\infty)}=\frac{\sqrt{-g}}{16\pi G} \left( F^{\mu r}+\frac{4\kappa}{3}\epsilon ^{r\mu\nu\rho\sigma} A_{\nu} F_{\rho\sigma}\right)\Big{|}_{r\to\infty}\,.
\ee
The charge density can be computed as the time component of the current and is
\begin{equation}
n = \frac{\sqrt{3} Q}{8 \pi G L^3}\,.
\end{equation}
It can be easily checked that this current is not conserved. Using the equation of motion for the gauge field in \erf{eq:EOMs}, the divergence of the anomalous current reads
\be
\nabla_{\mu} J^{\mu}=-\frac{\sqrt{-g} \,\kappa}{48\pi G} \epsilon^{r\mu\nu\rho\sigma} F_{\mu\nu} F_{\rho\sigma}\Big{|}_{r\to\infty}=\frac{\kappa}{6 \pi G} E_{\mu} B^{\mu}\,,
\ee
where $E_{\mu}$ and $B_{\mu}$ correspond to the electric and magnetic fields of the boundary theory. The result above differs from that of \cite{Son:2009tf} in a numerical factor that can be explained taking into account the different normalization of the gauge field on top of the fact that we are using different definitions of the current: whereas we define it through its source term in the action, the current in \cite{Son:2009tf} is defined as the subleading term in the near boundary expansion of the gauge field \footnote{\label{footJ} The precise relation is $J^\mu_{\mathrm{here}} = \frac 1 2 \left(J^\mu_{\mathrm{SonSurowka}} + \frac{4 \kappa}{3(16G\pi)}\epsilon^{\mu\nu\rho\lambda}A_\nu F_{\rho\lambda}\right)$ where our gauge fields and
Chern-Simons coupling differ by a factor of $\frac 1 2 $ from \cite{Son:2009tf}.}. Taking these differences into account we find that
our anomaly coefficient $c$ is $1/3$ of the anomaly in \cite{Son:2009tf}.

In an analogous way, we can work out the stress-energy tensor of the gauge theory from the bulk action to first order in perturbations of the metric. In this case, it is 
necessary to include the Gibbons-Hawking term in order to keep a well-defined variational principle and also a counterterm coming from regularization of the boundary 
action \cite{Balasubramanian:1999re},
\bea \label{eq:gravaction}
S_{GH}&=&\frac{1}{8\pi G} \int \dd ^4 x \sqrt{-\gamma} K \,,\nn\\
S^{gravity}_{CT}&=&\frac{1}{8\pi G} \int \dd ^4 x \sqrt{-\gamma} \left( \frac{3}{L}+\frac{L}{4}R_{(4)}\right)\,,
\eea 
where $\gamma$ is the induced metric on the AdS boundary, $K$ is the extrinsic curvature and $R_{(4)}$ is the Ricci scalar of the boundary metric. The 
Chern-Simons term does not depend on the metric since it is a topological contribution, hence the stress tensor will not receive any correction due to its presence. 
In fact, one recovers the result for the energy-momentum tensor obtained in \cite{Balasubramanian:1999re} from a generic five dimensional asymptotically AdS geometry.

The action to second order in perturbations of the metric and the gauge field will receive contributions from the bulk action \erf{eq:bulkaction}, from the 
Gibbons-Hawking term and from the gravity counterterm \erf{eq:gravaction}, but also from the counterterm coming from regularization of the gauge part of the action 
needed to cancel the logarithmic divergence. This term is given by
\be
S^{gauge}_{CT}=-\frac{L}{16\pi G} \log{r} \int \dd ^4 x \sqrt{-\gamma} F_{\mu\nu} F^{\mu\nu}\,.
\ee
The precise form of the contribution of each of these terms to the second order action can be found in Appendix \ref{sec:app2nd}.

Notice that the expansion of the action includes both terms to second order in gauge fluctuations and in metric fluctuations, but also mixed terms. In principle 
one could think that only the mixed terms will contribute to the off-diagonal correlators, but of course this is not true due to the holographic operator mixing 
under the renormalization group flow.  Thus {\it a priori} all the terms contribute to all the correlation functions. Collecting the perturbation fields $a_M$ 
and $h_{MN}$ in a single vector $\Phi_I$ and inserting its Fourier mode decomposition,
\be 
\Phi^I(r,x^{\mu})=\int \frac{\dd^d k}{(2\pi)^d} \Phi^I_k (r) \e^{-i \omega t+i \vec{k}\vec{x}} \,,
\ee
the complete second order action on-shell can be compactly written as a boundary term
\be \label{eq:2ndor}
\delta S^{(2)}=\int \frac{\dd^d k}{(2\pi)^d} \lbrace \Phi^I_{-k} \cA_{IJ} \Phi '^J_k + \Phi^I_{-k}  \cB_{IJ} \Phi^J_k \rbrace\Big{|}_{r\to\infty}\,,
\ee
where derivatives are taken with respect to the radial coordinate. In order to avoid double counting and to keep the correct contact terms in the Green functions 
one should be careful in the extraction of the contribution to the $\cB$ matrix coming from the bulk action. 

We can compute the holographic response functions from \erf{eq:2ndor} by applying the prescription of \cite{Son:2002sd,Herzog:2002pc}. As first noticed in 
\cite{Amado:2009ts}, for a coupled system the holographic computation of the correlators amounts to find a maximal set of linearly independent solutions that 
satisfy infalling boundary conditions on the horizon and that source a single operator at the AdS boundary. Following \cite{Kaminski:2009dh}, 
we can construct a matrix of solutions $F^I\,_J (p,r)$ such that each of its columns corresponds to one of the independent solutions and that at the boundary becomes 
the unit matrix. Therefore, given a set of boundary values for the perturbations, $\varphi^I_k$, the bulk solutions are 
\be\label{eq:f}
\Phi^I_k (r) = F^I\,_J (k,r)\, \varphi^J_k\,.
\ee
The F matrix actually corresponds to the bulk-to-boundary propagator matrix for a coupled system of bulk fields, for which the dual field 
theory operators are well-defined at the UV scale and get mixed under RG flow. Inserting these solutions in the second order boundary action, the holographic 
Green functions are finally given by
\be\label{eq:GR}
G_{IJ}(k)= -2 \lim_{r\to\infty} \left(\cA_{IM} (F^M\,_J (k,r))' +\cB_{IJ}\right)\,.
\ee

Let us now present the precise setup that we will consider from now on in order to study the anomaly effects. Without loss of generality we consider perturbations 
of frequency $\omega$ and momentum $k$ in the $y$-direction. We discuss only the shear channel 
(transverse momentum propagation) since the off-diagonal correlators appearing in the Kubo formulas for the chiral vortical and the chiral magnetic transport 
coefficients belong to the vector channel. This implies that 
we have to switch on the fluctuations $A_i$, $h^i_{\,t}$ and $h^i_{\,y}$, where $i=x,z$. For convenience we define new parameters
\be
a=\frac{\mu^2 L^2}{3\, r_{\rm H}^2}\quad;\qquad b= \frac{L^2}{2\, r_{\rm H}}\,,
\ee
and the compact coordinate $u=r_{\rm H}^2/r^2$, for which the horizon sits at $u=1$ and the AdS boundary at $u=0$. 

Finally we can write the system of differential equations for the shear sector, that consists on six second order equations and two constraints
\bea
\label{constraint}\hspace{-0.65cm} 0 &=& \omega h'^i_t(u) + k\, f(u)\, h'^i_x(u) - 3\,a\,\omega \,u\, B_i(u)\,,\\
\label{htx}\hspace{-0.65cm} 0&=&  h''^i_t(u)-\frac{h'^i_t(u)}{u}-\frac{b^2\,(k^2\,h^i_t(u)+k\,\omega\, h^i_x(u))}{u\,f(u)}-3\,a\,u\,B'_i(u)\,,\\ 
\label{hxy}\hspace{-0.65cm} 0 &=&  h''^i_x(u)+\left(\frac{f'(u)}{f(u)}-\frac{1}{u}\right)\,h'^i_x(u)-\frac{b^2}{u\,f(u)^2}(k\,\omega\, h^i_t(u)+\omega^2\,h^i_x(u))\,,\\
\label{bx}\hspace{-0.65cm} 0&=&  B''_i(u)+\frac{f'(u)}{f(u)}B'_i(u)-\frac{b^2(\omega^2-k^2\,f(u))}{u\,f(u)^2}\,B_i(u)-i\,\epsilon_{ij}\,\bar\kappa \frac{k\, B_j(u)}{f(u)}-\frac{h'^i_t(u)}{f(u)}\,,
\eea
where we define $B_i(u)=A_i(u)/\mu$ and $\bar\kappa=4\kappa L^3\mu/r_H^2$. There still remains a residual gauge symmetry on the gravity sector $\delta h^i_t = \omega \lambda$ and  $\delta h^i_x = -k\lambda$. This symmetry tells us that the metric components are not independent. In fact, it is possible to combine \erf{constraint}  with \erf{htx} to obtain \erf{hxy}. 

Since we are interested in computing correlators at zero frequency, we can drop out the frequency dependent parts in the equations and solve the system up to first order in $k$. In this limit, the fields $h^i_x$ decouple from the system and take a constant value. The reduced system can be written as
\bea
\label{htx1}0&=&  h''^i_t(u)-\frac{h'^i_t(u)}{u}-3\,a\,u\,B'_i(u)\,,\\ 
\label{bx1} 0&=&  B''_i(u)+\frac{f'(u)}{f(u)}\,B'_i(u)-i\,\epsilon_{ij}\,\bar\kappa \frac{k\, B_j(u)}{f(u)}-\frac{h'^i_t(u)}{f(u)}\,,
\eea
where we expand the fields to first order in the dimensionless momentum $p=k/4\pi T$
\bea
h^i_t(u) &=& h^{(0)}_i(u)+p \,h^{(1)}_i(u) \,,\\
B_i(u) &=& B^{(0)}_i(u)+p \,B^{(1)}_i(u)\,.
\eea

In order for the solutions to be physically sensible, they have to satisfy certain boundary conditions. The first condition is that they source the desired single operators in the UV, so the bulk fields must satisfy  $h^i_t(0)=\tilde h^i_t$, $B_i(0)=\tilde B_i$, where the `tilde' parameters are the sources of the boundary operators.  
The second condition comes from imposing infalling boundary conditions at the horizon. But what is the `infalling' condition at zero frequency? At first sight, the condition leads to an ambiguity, since both solutions for the metric fluctuations are perfectly regular, allowing us to take an arbitrary horizon value. However, if we analyze the near horizon behavior of the fields for arbitrary frequency, 
\bea
h_t^i (u) &\sim& (1-u)^{- i\omega /4\pi T+1}\,,\\
B_i (u) &\sim& (1-u)^{-i\omega/4\pi T}\,,
\eea
and then take the limit $\omega\to 0$, we see that infalling in this case means: regular for the gauge field and vanishing for the metric fluctuation. In the appendix \ref{solw0} we present the solutions to the equations of motion at $\omega=0$ that satisfy this boundary problem. The apparent freedom in fixing that integration constant is related with the zero mode found for the fluid velocity in first order hydrodynamics. We devote Appendix \ref{vdiff0} to explore this relation and show that the transport coefficients $\xi_B$ and $\xi_V$ are independent of this parameter as one would expect.

Now that we have the solutions for the perturbations, we can go back to the formula \erf{eq:GR} and compute the corresponding holographic Green functions. 
If we consider the vector of fields to be 
\be
\Phi_k^{\top} (u) = \Big{(} B_x(u) ,\, h^x_{\,t}(u)  ,\, B_z(u) ,\, h^z_{\,t}(u) \Big{)} \,,
\ee
the $\cA$ and $\cB$ matrices for that setup take the following form

\be
\cA=\frac{r_{\rm H}^4}{16\pi G L^5} \,{\rm Diag}\left( -3\, a f,\, \frac{1}{u} ,\, -3\, a f,\, \frac{1}{u} \right) \,,
\ee

\bea
\hspace{-0.8cm}\cB_{AdS+\partial}=
\frac{r_{\rm H}^4}{16\pi G L^5}
\left(
\begin{array}{cccc}
0 & -3a  & \frac{4 \kappa i p \mu^2 \phi L^5}{3 r_{\rm H}^4} & 0  \\
0 & -\frac{3}{u^2} &  0 & 0  \\
\frac{-4 \kappa i p \mu^2 \phi L^5}{3 r_{\rm H}^4} & 0 & 0 & -3a  \\
0 & 0 & 0 & -\frac{3}{u^2}  \\
\end{array}
\right)\,,
\eea

\bea
\cB_{CT}=
\frac{r_{\rm H}^4}{16\pi G L^5}
\left(
\begin{array}{cccc}
\vspace{0.15cm}
0&  0 & 0 & 0 \\
\vspace{0.15cm}
0 & \frac{3}{u^2 \sqrt{f\,}} & 0 & 0  \\
\vspace{0.15cm}
0 & 0  & 0 & 0\\
\vspace{0.15cm}
0 & 0  & 0 & \frac{3}{u^2 \sqrt{f\,}}  \\
\end{array}
\right)\,,\nn \\
\hspace{-0.8cm}\,
\eea
where we have split the $\cB=\cB_{AdS+\partial}+\cB_{CT}$ matrix into the contribution coming from the bulk and the Gibbons-Hawking actions, $\cB_{AdS+\partial}$, and the contribution coming 
from the counterterms, $\cB_{CT}$. With these matrices and the perturbative solutions we can construct the matrix of propagators. The non-vanishing retarded correlation functions at zero frequency are then
\bea
\label{eq:giti}G_{x,tx} &=& G_{z,tz} =\frac{\sqrt{3}\, Q}{4 \pi\, G\, L^3  }\,, \\
\label{eq:gxz}G_{x,z} &=& - G_{z,x} =-\frac{i\, \sqrt{3}\, k\, Q\,  \kappa}{2 \pi\, G \, r_{\rm{H}}^2}-\frac{i\, k\, \beta \, \kappa}{6\pi\, G  }\,,\\
G_{x,tz} &=&G_{tx,z} = -G_{z,tx}=-G_{tz,x}=-\frac{3\, i\, k\, Q^2\,  \kappa }{4 \pi\,G\,  r_{\rm{H}}^4} \,,\\
G_{tx,tx} &=& G_{tz,tz}=\frac{M}{16\pi\, G\, L^3 }\,,\\
\label{eq:gtiti}G_{tx,tz} &=& -G_{tz,tx}=-\frac{i\, \sqrt{3}\, k\, Q^3\, \kappa}{2\pi\, G\, r_{\rm{H}}^6}\,.
\eea
The off-diagonal current-current correlator \erf{eq:gxz} is consistent with the axial-axial current correlator computed on \cite{Gynther:2010ed}. 
As argued in the introduction, a background gauge field corresponds to a deformation of the dynamics of the boundary theory. For an anomalous theory, this deformation 
would lead to the appearance of an extra Chern-Simons term in the current constitutive relation
\be\label{deltaJ}
\delta J^\mu=\frac{4\kappa}{3(16\pi G)}\epsilon^{\mu\nu\rho\lambda}A_\nu F_{\rho\lambda}\,,
\ee
which is not gauge invariant but is allowed because of the anomaly. If we deform the theory turning on  a  constant background $A_0=\beta$, we get a contribution to the magnetic conductivity coming from (\ref{deltaJ})  which is precisely the $\beta$-part on (\ref{eq:gxz}). This extra contribution to the constitutive relation is just the difference between the current defined as the subleading term in the boundary expansion of the gauge field and the definition we are using, i.e. as the variation of the action including the terms stemming from the Chern-Simons part. Therefore, a non-vanishing $\beta$ can be interpreted as a source for such a topological coupling. In fact, differentiating with respect to this background field it is possible to compute the three point function of currents
\be
\langle J^i(k)J^j(-k)J^t(0)\rangle=-\frac{i\, k \, \kappa}{6\pi\, G  }\epsilon_{ij}= - i\, k\, c\,\epsilon_{ij}\,.
\ee

On the other hand, if we switch off the deformation, i.e. vanishing background field, we recover the conductivities \erf{sigmab}, \erf{sigmav}, \erf{eq:cmc} and \erf{eq:cvc} 
\bea
\label{eq:sigb}\sigma_B &=& \frac{\sqrt{3}\,  Q \, \kappa}{2 \pi \,G\,  r_{\rm{H}}^2} =  (3 c)\, \mu\,,\\
\label{eq:sigv}\sigma_V &=&\frac{3 \,  Q^2 \, \kappa }{4 \pi\, G\,  r_{\rm{H}}^4} =  (3 c)\, \frac{\mu^2}{2}\,, \\
\label{eq:xib}\xi_B &=& \,\frac{ \sqrt{3} \, Q\, \left(M\, L^2 \,+\,3\, r_{\rm{H}}^4\right)\,\kappa}{8\pi\, G\, M L^2\,  r_{\rm{H}}^2} = 
 (3c)\, \left( \mu - \frac{1}{2} \frac{n \mu^2}{\epsilon+P}\right)\,,\\
\label{eq:xiv}\xi_V &=&\frac{3 \,Q^2 \,\kappa}{4 \pi\, G\, M\,L^2   } =  \frac{3c}{2}\, \left( \mu^2 - \frac{2}{3} \frac{n \mu^3}{\epsilon+P}\right)\,.
\eea
The last two expressions agree precisely with \cite{Son:2009tf} once we take into account the difference in the
anomaly factor $c$ and the fact that we also dropped a factor $1/2$ in our definition of the vorticity.
They also obey of course the relations (\ref{eq:integrali}), (\ref{eq:integralii}) and (\ref{sigmav}).
The result for $\sigma_B$ coincides with the previous holographic computations of \cite{Gynther:2010ed, Yee:2009vw} for the chiral magnetic effect in an axial anomalous theory, and therefore with the weak coupling field theoretical result 
of \cite{Kharzeev:2009pj}, whereas the chiral vortical conductivity $\sigma_V$ is a new result.

\subsection{Frequency dependence}

In order to study the frequency dependence of the chiral conductivities, we can use \erf{sigmab} and  \erf{sigmav} to define 
\bea
\sigma_B(\omega)  &=& \lim_{k_m\to0}\frac{-i}{k_m} \epsilon_{mij} \vev{ J^i J^{j}}\,,\\
\sigma_V(\omega) &=&\lim_{k_m\to0}\frac{-i}{k_m} \epsilon_{mij} \vev{ J^i T^{tj}}\,.
\eea

It is important to notice that these, and not the $\xi_V$ and $\xi_B$, are the relevant conductivities at finite frequency. The latter correspond to the conductivities measured in the local rest frame of the fluid, where one subtracts the contribution to the current due to the energy flux generated when we put the system in a background magnetic or vorticity field. But as we have seen, there is an ambiguity in the definition of the local rest frame: the fluid velocity is frequency and momentum dependent, one can just define it order by order in the hydrodynamic gradient expansion up to an arbitrary contribution. This automatically implies that the $\xi_V $ and $\xi_B $ are only meaningful in the zero frequency limit. On the other hand, the $\sigma_V$ and $\sigma_B$ conductivities are not subject to this problem: they capture the complete response of the system to the external magnetic fields. Therefore, it is sensible to define the frequency dependent chiral conductivities above, in an analogous way as for the A.C. electric conductivity.

To study that dependence holographically, we have to resort to numerics. The nature of the system allows us to integrate from the horizon out to the boundary, so we should fix boundary conditions at the first one, even though we would like to be free to fix the AdS boundary values of the fields, hence the operator sources. Imposing infalling boundary conditions, the fluctuations can be written as
\bea
h^i_t(u) &=& (1-u)^{-i w+1}\,H^i_t(u)\,,\\
h^i_x(u) &=& (1-u)^{-i w}\,H^i_x(u)\,,\\
B^i(u) &=& (1-u)^{-i w}\,b^i(u)\,,
\eea
where $w=\omega/4\pi T$. As we saw, the remaining gauge symmetry acting on the shear channel implies that $h^i_t$ and $h^i_x$ are not independent. So if we fix the horizon value of the $\{b^i,H^i_t\}$ fields, the constraints \erf{constraint} fixes
\be
H^i_x(1)=-\frac{3ia b^i(1)+ (i+w)H^i_t(1)}{(2-a)p}\,.
\ee
In order to find a maximal set of linearly independent solutions, we can construct four of them using linearly independent combinations of these horizon free parameters. In this way we construct the following independent horizon value vectors
\bea
\begin{array}{cccc}
\left(\begin{array}{c}
1\\
0\\
-\frac{3ia}{(2-a)p}\\
0\\
0\\
0
\end{array}\right) ,&
\left(\begin{array}{c}
 0\\
 1\\
 -\frac{i+w}{(2-a)p}\\
 0\\
 0\\
 0
 \end{array}\right), &
\left(\begin{array}{c}
 0\\
 0\\
 0\\
 1\\
 0\\
 -\frac{3ia}{(2-a)p}\\
 \end{array}\right), &
\left(\begin{array}{c}
 0\\
 0\\
 0\\
 0\\
 1\\
 -\frac{i+w}{(2-a)p} 
 \end{array}\right).
 \end{array}
 \eea
The remaining two are given by pure gauge solutions arising from gauge transformations of the trivial one. We choose them to be
\bea
\Phi(u)=\begin{array}{cccc}
\left(\begin{array}{c}
0\\
w\\
-p\\
0\\
0\\
0
\end{array}\right) \,,&
\left(\begin{array}{c}
 0\\
 0\\
 0\\
 0\\
 w\\
 -p
 \end{array}\right)\,.
 \end{array}
 \eea
 Using the corresponding solutions we construct the $F$ matrix of \erf{eq:f} in this way:
 \be
 F^I_J(u) = H^I_M(u) H^{-1M}_J(0)\,,
 \ee
where $H^I_J(u)=(\Phi^I(u))_J$. 

\begin{figure}[!tp]
\begin{center}
\includegraphics[scale=0.58]{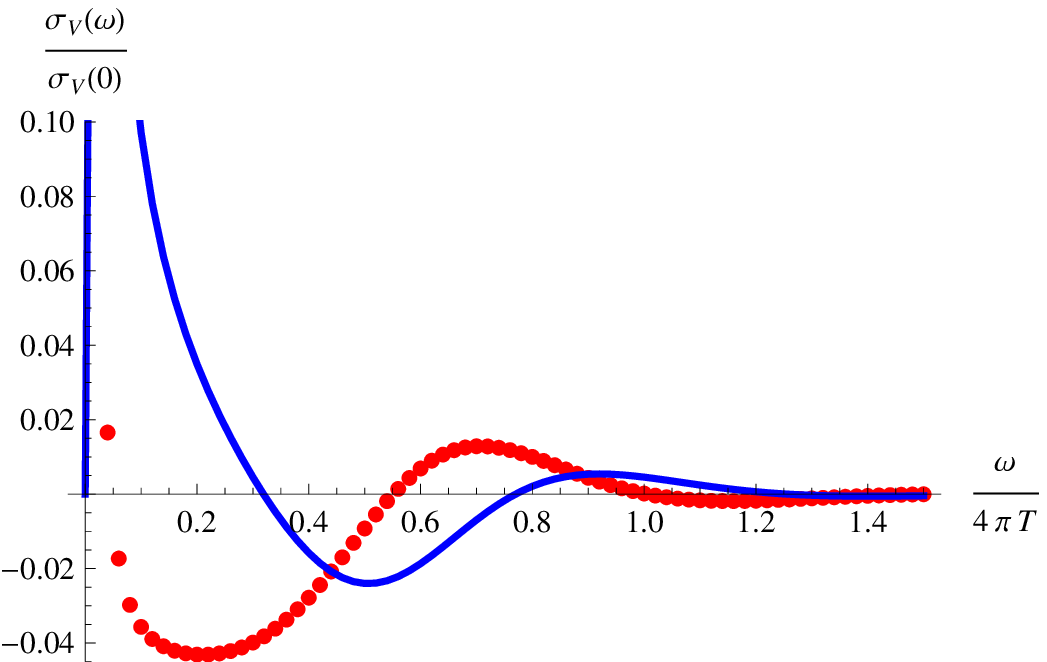}\hfill \includegraphics[scale=0.58]{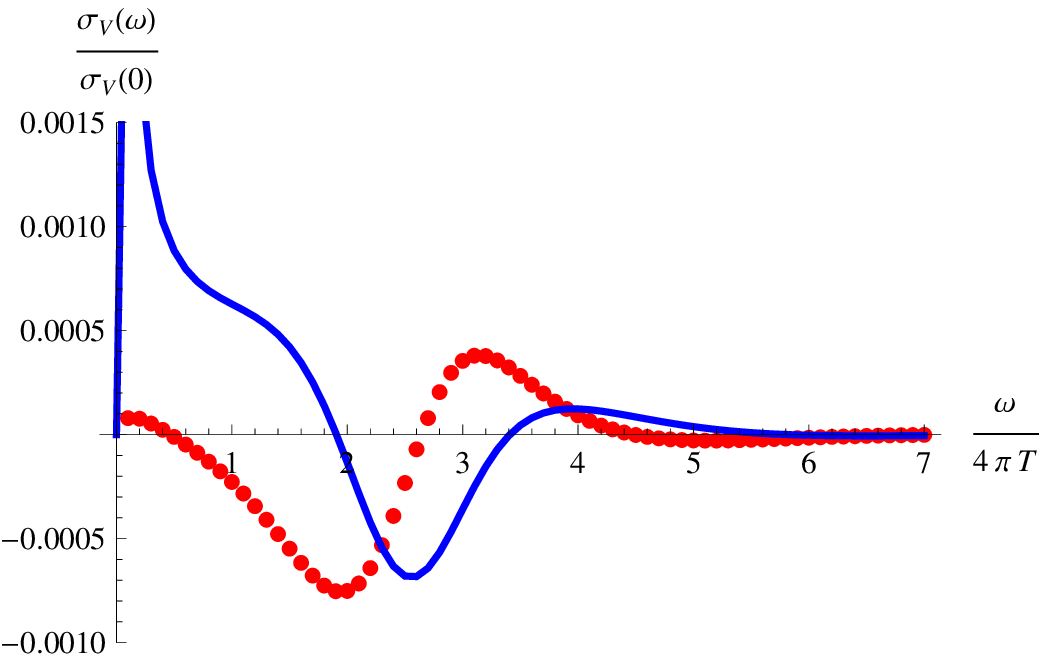} \\
\vspace{0.23cm}
\includegraphics[scale=0.58]{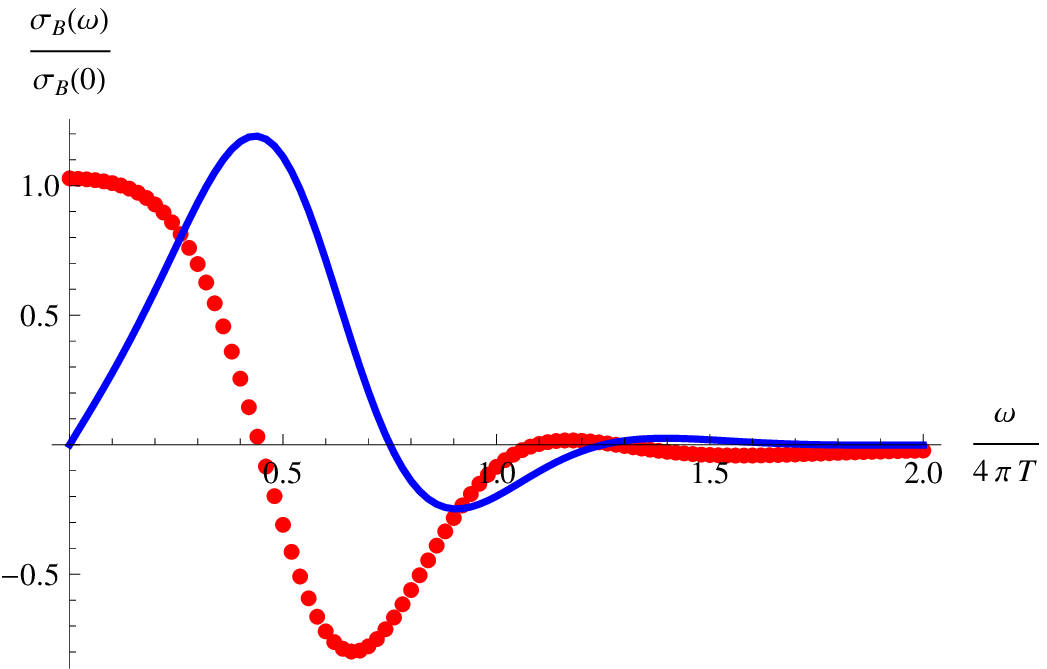} \hfill\includegraphics[scale=0.58]{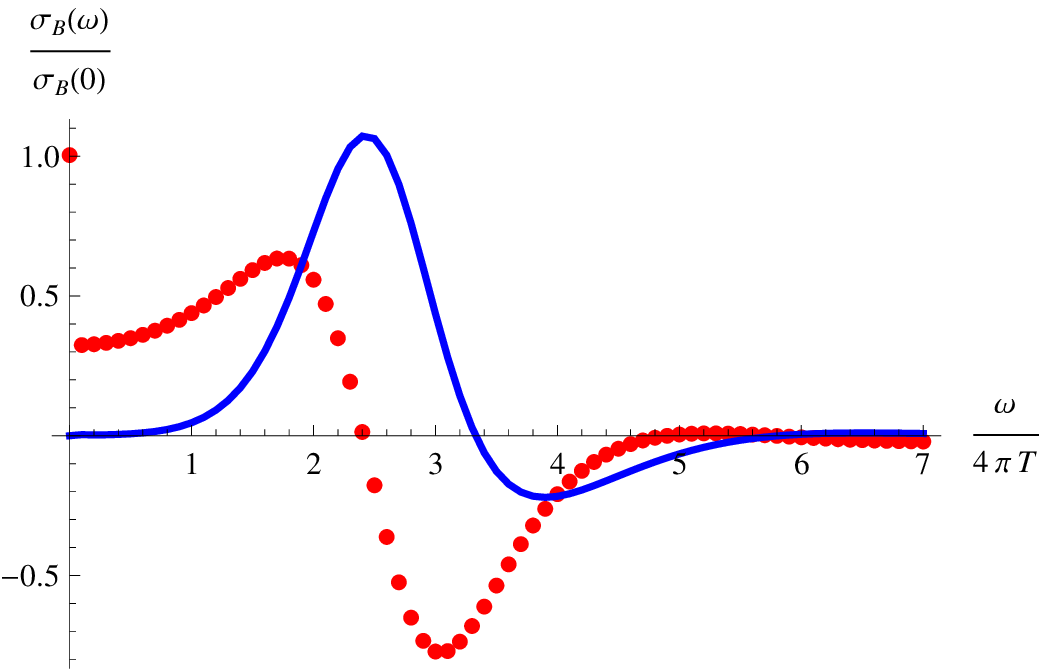} 
\end{center}
\caption{\label{plots1}Chiral vortical (up) and magnetic (bottom) conductivities as function of the frequency at $\tau=36.5$ (left) and $\tau=0.24$ (right). Red doted points represent real part and thick blue line the imaginary conductivity.}
\end{figure}

\begin{figure}[!tp]
\begin{center}
\includegraphics[scale=0.58]{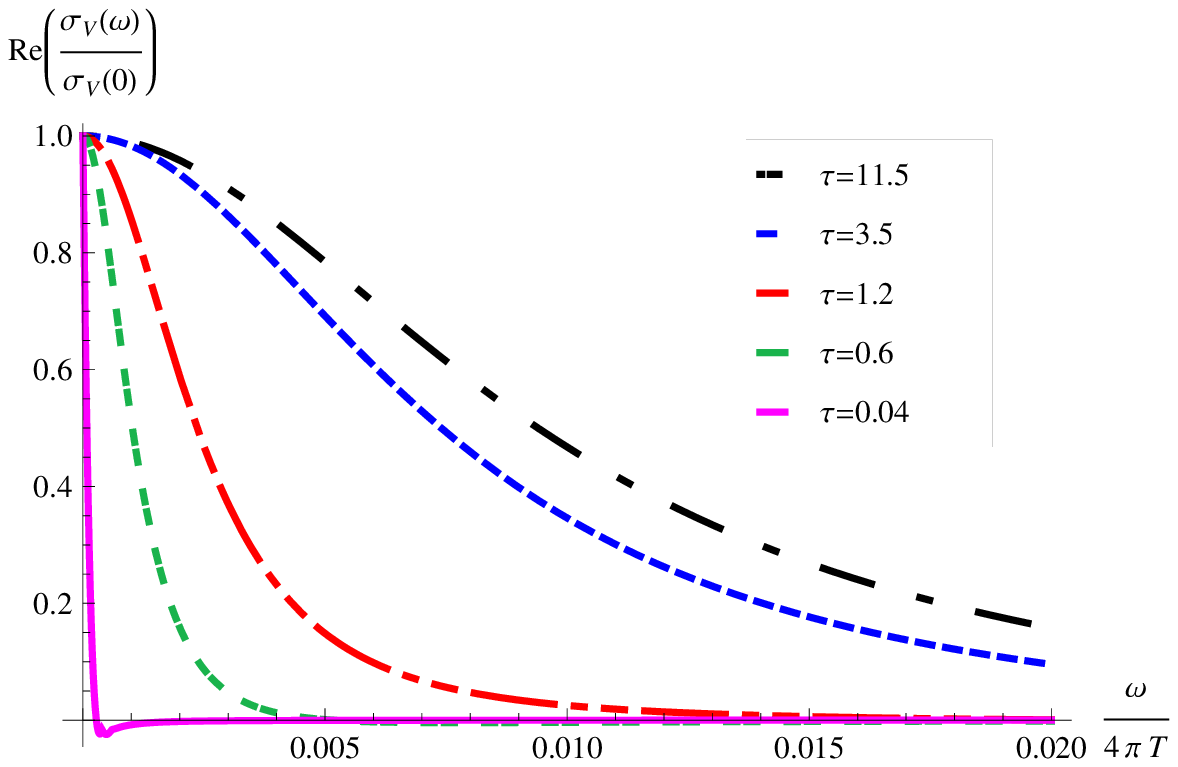}\hfill \includegraphics[scale=0.58]{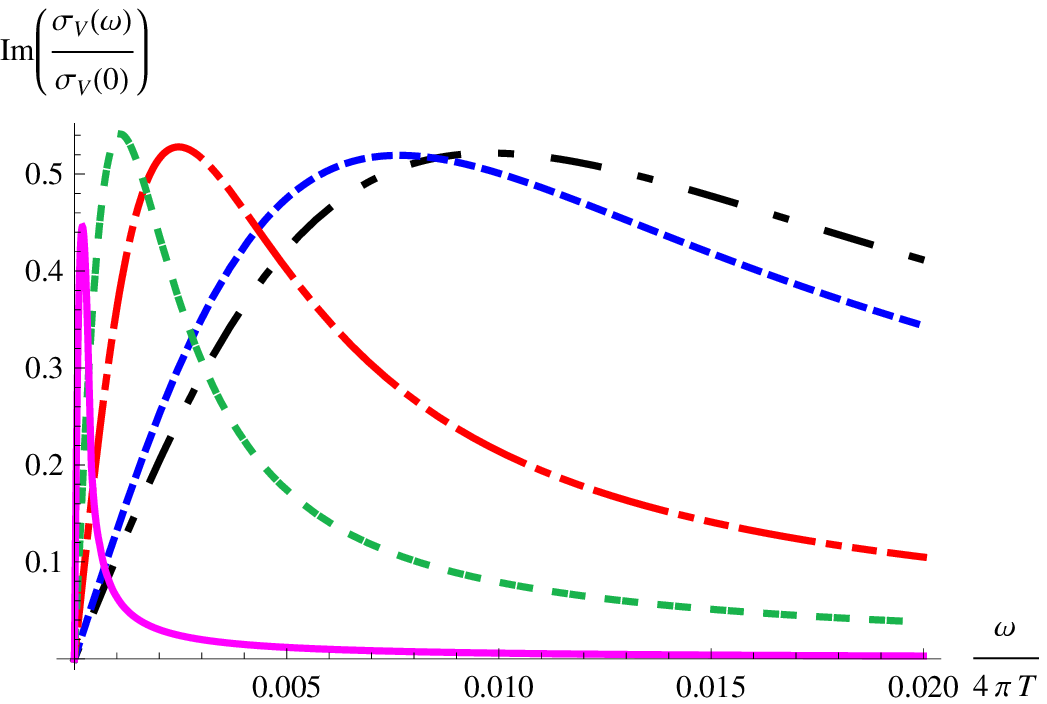} \\
\includegraphics[scale=0.58]{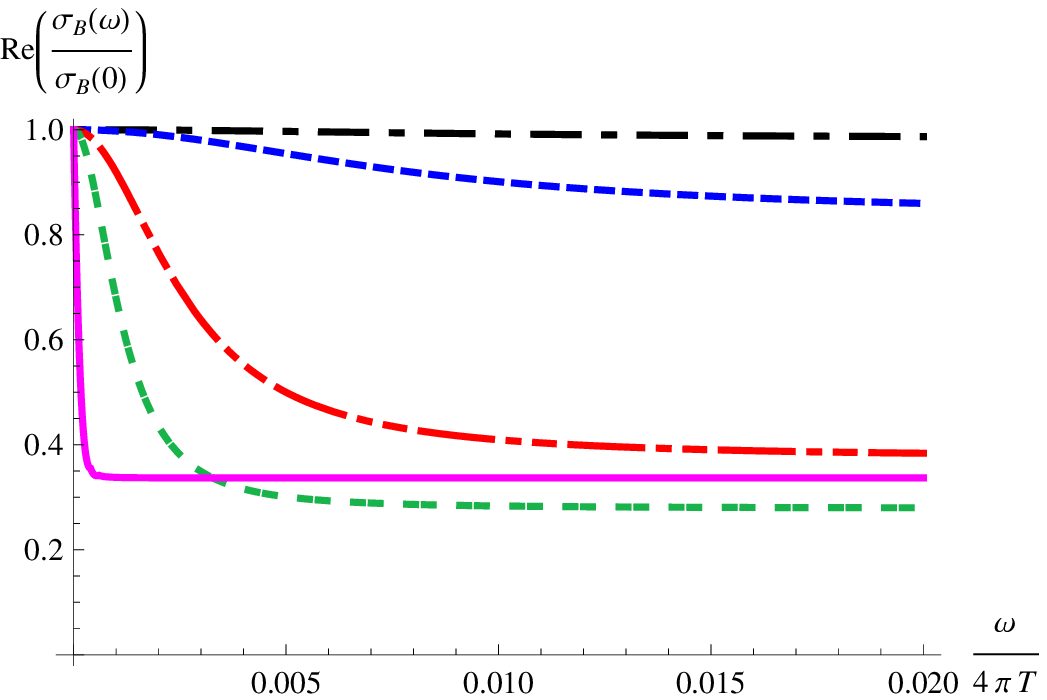} \hfill\includegraphics[scale=0.58]{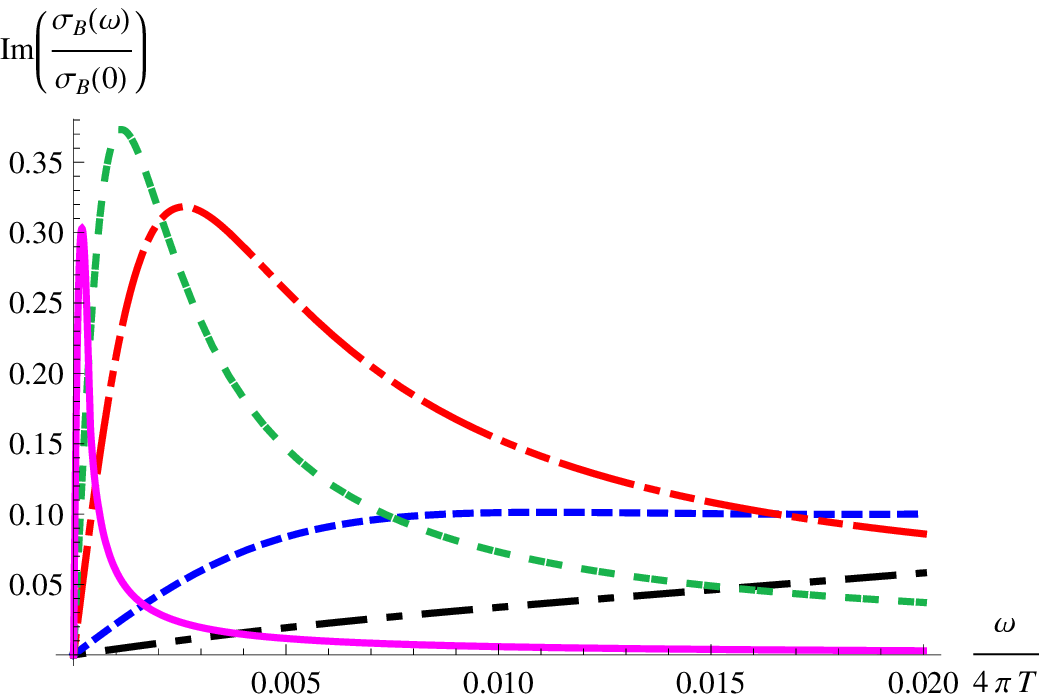} 
\end{center}
\caption{\label{plots2}Chiral vortical (up) and magnetic (bottom) conductivities as function of the frequency close to $\omega=0$. Real (left) and imaginary (right) part of the normalized conductivity for different values of the dimensionless temperature.}
\end{figure}

In Figure \ref{plots1} is illustrated the behavior of the vortical and magnetic conductivities as a function of frequency for two very different values of the dimensionless temperature $\tau=2\pi r_{\rm H}T/\mu$. Both of them go to 
their corresponding zero frequency analytic result in the $\omega\to0$ limit. The frequency dependent chiral magnetic conductivity was also computed in \cite{Yee:2009vw}, 
though in that case the possible contributions coming from metric fluctuations were neglected. Our result for $\sigma_B(\omega)$ agrees pretty well with the result found in that work in the case of high temperature when the metrics fluctuations can be neglected, but it develops 
a dip close to $\omega=0$ when temperature is decreased (see Figure \ref{plots2}), due to the energy flow effect. For small temperatures, the chiral magnetic conductivity drops to $\sim1/3$ of its zero frequency value as soon as we move to finite frequency. 
The behavior of $\sigma_V$ is slightly different: the damping is much faster and the imaginary part remains small compared with the zero frequency value.

\begin{figure}[!htp]
\begin{center}
\includegraphics[scale=0.58]{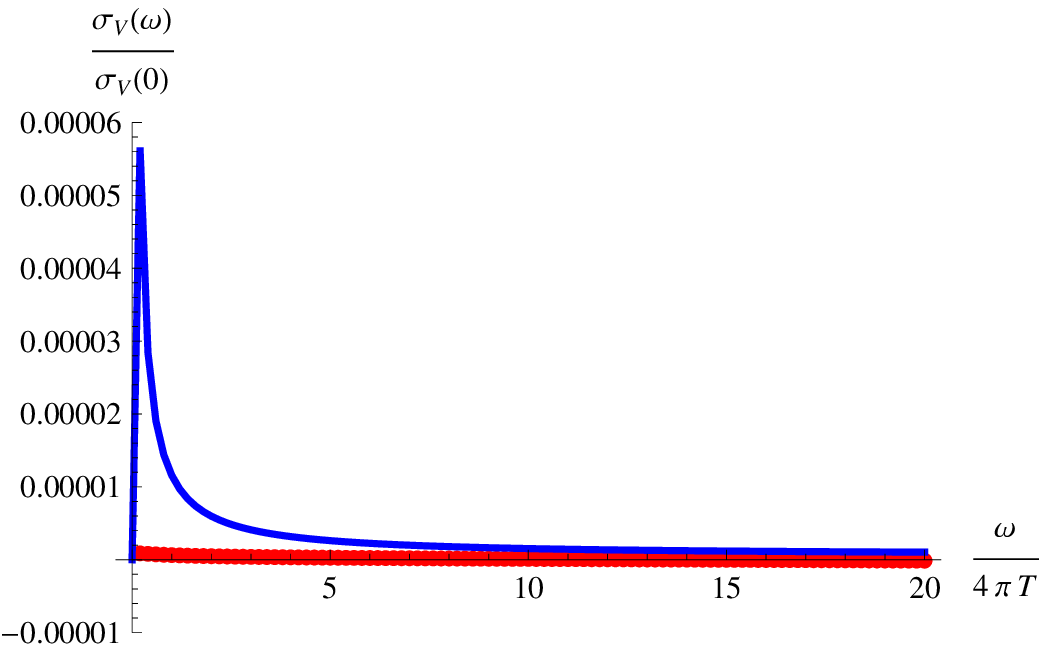}\hfill \includegraphics[scale=0.58]{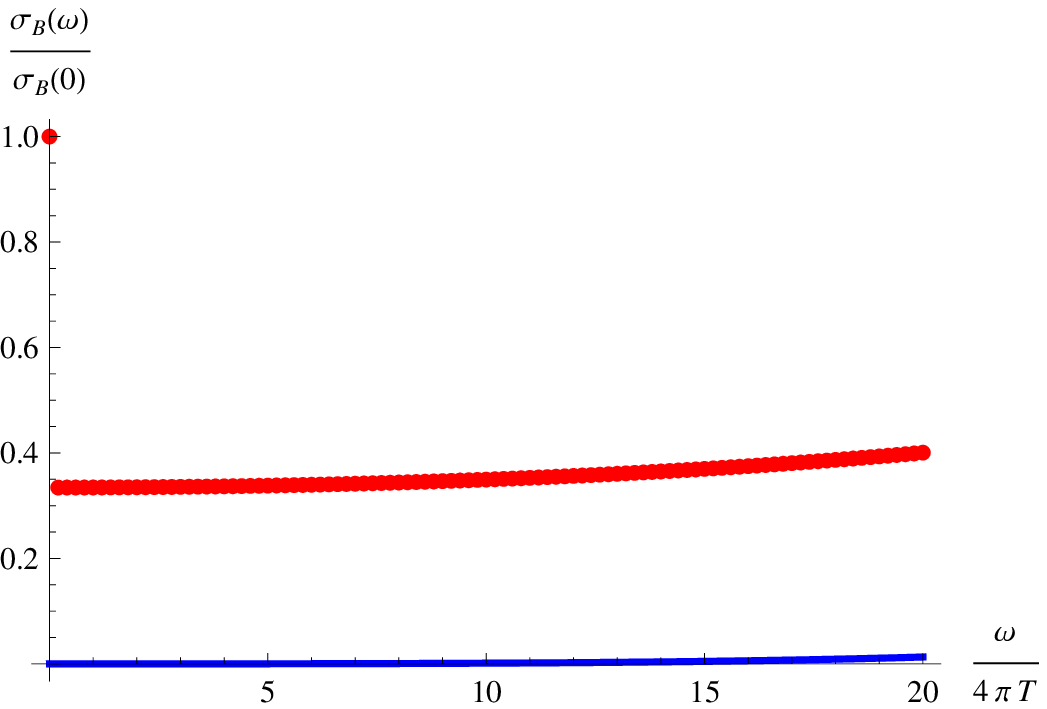}
\end{center}
\caption{\label{plots3}Chiral vortical (left) and magnetic (right) conductivities as function of the frequency at $\tau=0.008$.   Red doted points represent real part and thick blue line the imaginary conductivity.  The real part of $\sigma_V$ at $\omega=0$ is outside the range of the plot.}
\end{figure}

In Figure \ref{plots2} we made a zoom to smaller frequencies in order to see the structure of the dip on $\sigma_B$ and the faster damping on $\sigma_V$. In Figure \ref{plots3} we show the conductivities for very small temperature. From this plots we can infer that at zero temperature the conductivities behave like $\sigma_B = \alpha\sigma_B^0\left(1+\frac{1-\alpha}{\alpha}\delta(\omega)\right)$ and $\sigma_V=\sigma_V^0\delta(\omega)$ with $\alpha$ a constant value of order $1/3$.

%% file: conclusion.tex
\section{Conclusion and Outlook}

We have derived Kubo formulas for the anomalous magnetic and vortical conductivities $\xi_B$, $\xi_V$ as
they are defined through the constitutive relations in the 
Landau frame for an anomalous current. These Kubo formulas can be used in order to perform a first principle field theoretical computation of the chiral vortical effect in a 
weakly coupled anomalous theory in an analogous way as it has been done for the chiral magnetic effect in \cite{Kharzeev:2009pj}. Such result would allow us to 
confirm that the chiral vortical effect is coupling independent, as it is expected since it is anomaly induced.

The peculiar form of the Kubo formula relies on the fact that these coefficients are defined for the local rest frame, so they measure the current generated in the presence of external background fields after subtraction 
of the energy flow contribution. This flow contribution is fixed once we have chosen a frame (e.g. Landau frame) for the definition of the fluid velocity. For this reason we have also defined elementary conductivities, $\sigma_B$ and $\sigma_V$, that measure the complete response to the 
external magnetic and gravitomagnetic field. The analytic results found for the zero frequency conductivities were given in \erf{eq:sigb} -- \erf{eq:xiv}. They reproduce the known holographic results 
for the $\xi$ conductivities. The chiral magnetic conductivity $\sigma_B$ also reproduces the weak coupling results for an axial anomalous theory, whereas the chiral vortical conductivity $\sigma_V$ is 
a genuine new result. The frequency dependence of the $\sigma_B$ shows that at very small temperatures, the conductivity drops to $1/3$ of its zero frequency value as soon as one moves to $\omega\neq0$. On the other hand, 
even for large temperature, $\sigma_V$ decays very fast as a function of the frequency. It would be interesting to have a weak coupling computation of the chiral vortical conductivity in order to see if 
the strong interactions have some effect\cite{progress}. In principle, both effects are experimentally accessible since they induce a macroscopical effect, they modify the hydrodynamic description of the system. In fact, they have been argued for being relevant in heavy ion collisions to explain the charge asymmetry fluctuations observed at RHIC for non-central collisions.

%% file: second-order-sol.tex
\section{Second Order Action}\label{sec:app2nd}
The general contributions to the second order action in perturbations of the gauge field and the metric coming from the bulk action, the Gibbons-Hawking term and 
the gravity and gauge counterterms, respectively, are
\bea
\hspace{-0.3cm}\delta S^{(2)}&=&\frac{1}{16\pi G} \int \dd ^4 x\sqrt{-g} \Big{\lbrace} -\left( \frac{4 \kappa}{3} \epsilon ^{rMABC} A_M+\frac{1}{2}\left( g^{Br} g^{AC}- g^{Cr} g^{AB}\right)\right) a_A \nabla _B a_C\nn\\
&+& \Big{(}\frac{3}{4}\left(g^{AD}g^{BE}g^{rC}+g^{AC}g^{DE}g^{rB}\right)+\frac{1}{4} g^{AB}\left(g^{CD}g^{rE}-g^{DE}g^{rC}\right) \nn\\
 &-&g^{AD}g^{CE}g^{rB}-\frac{1}{2}g^{AD}g^{BC}g^{rE} \Big{)} h_{AB} \nabla _C h_{DE} \nn\\
&+&\frac{1}{4}\left( g^{AB} F^{rC}+2\left( g^{Ar} F^{BC}-g^{AC} F^{Br}\right)\right) h_{AB} a_C\Big{\rbrace} \Big{|}_{r\to\infty} \,,
\eea
\bea
\hspace{-0.3cm}\delta S_{GH}^{(2)}&=&-\frac{1}{16\pi G} \int \dd ^4 x\sqrt{-\gamma} \Big{\lbrace} \frac{1}{2} \left(n^{C}g^{AB}g^{DE}-2 n^{C}g^{AD}g^{BE}\right) h_{AB}\nabla _C h_{DE} \\
\hspace{-0.3cm}&+&\frac{1}{4} \left(g^{AB} g^{CD}-2 g^{AC} g^{BD}\right) \nabla _E n^E \,h_{AB} h_{CD}- \nabla _A \left(h h^{AB}n_B -2 h^{AC}h_C\,^B n_B\right)\Big{\rbrace} , \nn
\eea
where $n^A$ is the normal vector to the boundary.
\bea
\hspace{-0.3cm}\delta S_{CT}^{grav (2)}&=&\frac{1}{16\pi G} \int \dd ^4 x\sqrt{-\gamma} \Big{\lbrace} \frac{1}{4} \left( g^{\mu\nu}g^{\rho\sigma}-2g^{\mu\rho}g^{\nu\sigma}\right) \left(\frac{3}{L}+\frac{L}{4}R_{(4)}\right) h_{\mu\nu} h_{\rho\sigma} \nn\\
&& \hspace{-1.5cm} -\,\frac{L}{4}\left(g^{\mu\nu} R_{(4)}^{\rho\sigma}-2g^{\nu\rho} R_{(4)}^{\mu\sigma}\right)h_{\mu\nu} h_{\rho\sigma}+\frac{L}{4} \Big{(}\frac{1}{2} h(\nabla _{\mu} \nabla _{\nu} h^{\mu\nu}-\nabla _{\mu} \nabla ^{\mu} h)+h^{\mu\nu}\nabla _{\mu} \nabla _{\nu} h \nn\\
&& \hspace{-1.5cm} +\,h^{\mu\nu}\nabla _{\alpha} \nabla ^{\alpha} h_{\mu\nu}- h^{\mu\nu}\nabla _{\mu} \nabla _{\alpha} h_{\nu}\,^{\alpha}-h^{\mu\nu}\nabla _{\alpha} \nabla _{\mu} h_{\nu}\,^{\alpha} +\nabla_{\mu} h^{\mu\nu} \nabla_{\nu} h-\frac{1}{4} \nabla_{\mu} h \nabla^{\mu} h \nn\\
&& \hspace{-1.5cm} +\,\frac{3}{4} \nabla_{\alpha} h_{\mu\nu} \nabla^{\alpha} h^{\mu\nu}-\nabla_{\mu} h^{\mu\nu} \nabla_{\alpha} h_{\nu}\,^{\alpha}-\frac{1}{2} \nabla_{\alpha} h^{\mu\nu} \nabla_{\mu} h_{\nu}\,^{\alpha}\Big{)}\Big{\rbrace} \,,
\eea 
\bea
\hspace{-0.3cm}\delta S_{CT}^{gauge (2)}&=& -\frac{L}{16\pi G} \log{r} \int \dd ^4 x \sqrt{-\gamma} \Big{\lbrace} \frac{1}{8} F^2\left( h^2- 2 h_{\mu\nu}h^{\mu\nu} \right) +2 F_{\mu\nu} F^{\beta\nu} h^{\mu\alpha}h_{\alpha\beta} \nn\\
&&\hspace{-0.8cm}+\, F_{\mu\nu} F_{\alpha\beta} h^{\mu\alpha}h^{\nu\beta}-F_{\mu}\,^{\nu}F_{\alpha\nu} h h^{\mu\alpha}+2\left(g^{\mu\alpha}g^{\nu\beta}-g^{\mu\beta}g^{\nu\alpha}\right) \nabla_{\mu} a_{\nu} \nabla_{\alpha} a_{\beta}\nn\\
&&\hspace{-0.8cm}+\,2 h F^{\mu\nu}\nabla_{\mu}a_{\nu}-4 h^{\mu\alpha} F_{\alpha}\,^{\nu} \nabla_{\mu}a_{\nu}\Big{\rbrace}\,.
\eea
From these terms we extract the $\cA$ and $\cB$ matrices in the boundary term \erf{eq:2ndor}.

%% file: anal-sol.tex
\section{Analytic solutions in the hydrodynamic regime}\label{solw0}
The perturbative solutions of the system \erf{htx1} and \erf{bx1} up to first order in momentum are 
\begin{align}
\nonumber h^i_t(u)&=\tilde h^i_t f(u)-\frac{i k \bar\kappa\epsilon_{ij} (u-1)a }{2 (1+4 a)^{3/2}}\left[3 \left(\sqrt{1+4 a} u (2 a u-1)+\right.\right.\\
&\left.2 \left(1+u-a u^2\right) \text{ArcCoth}\left[\frac{2+u}{\sqrt{1+4 a} u}\right]\right)\tilde B_j+
\left. (1+4 a)^{3/2} u^2 \tilde h^j_t\right]\,,\\
\nonumber B_i(u)&= \tilde B_i+\tilde h^i_t u -i\frac{k\bar\kappa\epsilon_{ij}}{2 (1+4 a)^{3/2}} \left(\tilde h^j_t u\left(  1+4 a   \right)^{3/2}+\right.\\
&\left.\tilde B_j \left(6 a \sqrt{1+4 a}  u+ (-2+a (-2+3 u)) \text{Log}\left[\frac{2+u-\sqrt{1+4 a} u}{2+u+\sqrt{1+4 a} u}\right]\right)\right)\,.
\end{align}

%% file: v-diff-0.tex
\section{Fluid velocity dependence}\label{vdiff0}

We have seen that in the hydrodynamic regime the velocity of the fluid in the Landau frame is determined modulo a $P$-odd term $v_m\sim O(k)$ that is an arbitrary function of the sources. In this 
appendix we show the independence of the transport coefficients on this arbitrary function, even if the correlators are velocity dependent, and also that these arbitrariness disappears once we correctly impose the physical boundary conditions on the bulk fields.

In order to do so, we are going to solve the system at $\omega=0$, first order in $k$ and for arbitrary value of $v_m$. Again, the system reduces to:
 \bea
 0&=&  h''^i_t(u)-\frac{h'^i_t(u)}{u}-3\, a\, u\, B'_i(u)\,,\\ 
0&=&  B_i(u)+\frac{f'(u)}{f(u)}\,B'_i(u)-i\,\epsilon_{ij}\,\bar\kappa\,\frac{k B_j(u)}{f(u)}-\frac{h'^i_t(u)}{f(u)}\,,
\eea
where $h^i_t(u) = h^{(0)}_i(u)+p \,h^{(1)}_i(u) $ and $B_i(u) = B^{(0)}_i(u)+p \,B^{(1)}_i(u)$. After imposing regularity at the horizon we find the following solutions:
\bea
B^{(0)}_i(u) &=& \tilde{B}_i + A_i\, u\,,\\\vspace{0.3cm}
B^{(1)}_i(u) &=& C_i u-\frac{2 i (1+a)^2 \bar\kappa \epsilon_{ij} A_j u}{(2-a) (1+4 a) b}-\frac{i\,\bar\kappa\, \epsilon_{ij} \tilde{B}_j}{(2-a)(1+4a)^{3/2}b}\left(9a (1+a) u \sqrt{1+4 a}\right.\\\vspace{0.2cm}
&&\hspace{-1.8cm}\left. +(2-a)^2 \left( 2 (1+a) \text{ArcCoth}\left[\sqrt{1+4 a}\right]+(2+a(2-3u)) \text{ArcTanh}\left[\frac{-1+2 a u}{\sqrt{1+4 a}}\right]\right)\right)\nn\\\vspace{0.3cm}
h^{(0)}_i(u) &=& \tilde{h}_i + A_i\,(f(u)-1)\,,\\\vspace{0.3cm}
h^{(1)}_i(u) &=&  C_i (f(u)-1) - \frac{i\, a\, \bar\kappa (4 (1+a)^2 u -27 a)\,\epsilon_{ij} A_j\, u^2}{2 (2-a) (1+4a) b} \\\vspace{0.2cm}
&&\hspace{-1.8cm}+\frac{3\,i\,a\,\bar\kappa\, \epsilon_{ij} \tilde{B}_j}{2(2-a)(1+4a)^{3/2}b} \left(((2+a(16+5a))u-6 a (1+a) u^2-(2-a)^2) \sqrt{1+4 a} u\right. \nn\\\vspace{0.2cm}
&&\hspace{-1.8cm}\left.+ 2 (2-a)^2 \left(\text{ArcCoth}\left[\sqrt{1+4 a}\right]+f(u) \text{ArcTanh}\left[\frac{-1+2 a u}{\sqrt{1+4 a}}\right]\right)\right)\,.\nn
\eea
As we know, this is not enough to solve the boundary value problem since both of the two independent solutions for the metric fluctuations satisfy the regularity condition. However, we can 
use the constitutive relations to try to fix the arbitrariness. In the hydrodynamic description, the stress-energy tensor is given by
\be
T^{ti}=(\epsilon +P) v^i -P \tilde{h}^{ti}\,,
\ee
where the velocity is order $p$. Using the holographic dictionary, we can identify the coefficient of the non-normalizable mode of the asymptotic behavior of a bulk field with the source of the dual operator and the coefficient 
of the normalizable one with its expectation value. Therefore, we can write the metric fluctuation close to the boundary as
\be
h_t^i(u)\sim \tilde{h}_t^i + T_t^i u^2\,,
\ee
so using the hydrodynamic result, we can do the identification order by order in momentum, in such a way that the 
velocity piece of the energy tensor fixes the horizon value of $h^{(1)}_i$. Doing so, the asymptotic behavior of each order becomes
\bea
h^{(0)}_i (u) &\sim& \tilde{h}_t^i (1-P u^2)\,,\nn\\
p h^{(1)}_i (u) &\sim& -(\epsilon +P) v_i u^2\,.
\eea 
We can proceed to construct the matrix of correlators for arbitrary value of the velocity as explained in section \ref{sec:holo}. Now, all the correlators pick contributions proportional to the 
velocity. In a compact way, the retarded propagators read
\bea
G_{i,j}&=&-\frac{r_{\rm H}}{\pi G L}\left(\frac{i\sqrt{3a} \,k\, (4+a) \kappa}{8 (1+a)}\epsilon_{ij}-\frac{r_{\rm H}}{2 L^2} \frac{\partial v_i}{\partial \tilde{B}_j}\right)-\frac{i k \beta \kappa}{6 \pi G} \epsilon_{ij}\,,\\
G_{i,tj}&=&-\frac{r_{\rm H}^2}{\pi G L^2}\left(\frac{3\, i\, a\, k\, \kappa}{4 (1+a)} \epsilon_{ij} -\frac{\sqrt{3a} r_{\rm H}}{4 L} \delta_{ij}-\frac{\sqrt{3a}\,r_{\rm H}\,}{2 L^2}\frac{\partial v_i}{\partial \tilde{h}_j}\right)\,,\\
G_{ti,j}&=&\frac{r_{\rm H}^3}{\pi G L^4} \frac{(1+a)}{\sqrt{3a}}\frac{\partial v_i}{\partial \tilde{B}_j}\,,\\
G_{ti,tj}&=&\frac{r_{\rm H}^4}{\pi G L^5 } \left(\frac{(1+a)}{16} \delta_{ij} + (1+a)\frac{\partial v_i}{\partial \tilde{h}_j}\right)\,,
\eea
where $i,j=x,z$. It is straightforward to prove that applying definitions \erf{eq:cvc} and \erf{eq:cmc}
for the chiral vortical and magnetic conductivities, the result is independent of the velocity and coincides with \erf{eq:xiv} and \erf{eq:xib} as expected. Setting the velocities 
to zero, the correlators coincide with those presented in \cite{Matsuo:2009xn}.

If we now impose the correct zero frequency `infalling' condition to the fields $h_t^i$, i. e. vanishing at the horizon, the velocities are not arbitrary anymore, but 
are given in terms of the boundary sources,
\be
v_i=-\frac{i a \bar\kappa \epsilon_{ij}(2 \tilde{h}_j + 3 \tilde{B}_j) k}{16(1+a)}\,.
\ee
Of course, substituting them in the Green functions given above, the antisymmetric correlation matrix spanned by \erf{eq:giti} -- \erf{eq:gtiti} is recovered.